\documentclass[a4paper,10pt]{article}
\usepackage{a4}
\begin{document}
\title{Around Poincare duality in discrete spaces}
\author{Alejandro Rivero\thanks{ 
Email: \tt rivero@wigner.unizar.es}}
\maketitle

\newcommand{\C}{{\bf C}}
\begin{abstract}

We walk out the landscape of K-theoretic Poincare Duality
for finite algebras, which will pave the way to get continuum Dirac
operators from discrete noncommutative manifolds.

\end{abstract}

\section{Generic}

This paper is a sequel to \cite{previous}\footnote{so please print it and 
staple both in the same folder...}. There we saw how a family of discrete
noncommutative spaces, namely those with intersection matrix
$$
q^{(n)}_{ij}=\pmatrix{  -1 & 1 & & 0 & 1  \cr 
              1 & -1 & 1 & & 0  \cr 
           & 1 & \ddots_n & 1 &   \cr 
              0 & & 1 & -1 & 1 \cr 
              1 & 0 & & 1 & -1 \cr}
$$
can be arranged to get a one dimensional commutative space, the circle
$S^1$, in the limit $n\to\infty$.

But also we noticed, just solving for the null eigenvectors of $q_{ij}$, that
the intersection matrix is degenerated for size $n$ multiple of six, then
putting in question Poincare Duality. 

Here we will examine some answers to this small nuisance.

The first one, obviously, is to reject such sizes. We have still subsequences
going to infinity with non degenerate $q$, so we can build the limit without
the multiples of 6. This was the approach of the previous paper, but one would
like to get a generic procedure, instead of a case-by-case approach.

For the same reason, one worries about the next easier procedure: modify the
intersection matrix to get a nondegenerate product while keeping the grading
sign. 
One possibility is to use an simpler spectral triple, at the cost
of losing the spatial homogeneity of this one, and even here we need to
control possible degenerated forms.
It is safer to increase the dimension of the Hilbert
spaces in the diagonal, $H_{ii}$, to 3 or bigger dimension, so the diagonal
elements change from $-1$ to $-3$, and Poincare duality works. Again, this
method is not generic enough, but it is interesting because it forces a
increase in the number of particles, just as in happens in Connes-Lot models.
On other hand, it sounds strange that even if $m_{ij}$ is diagonal and the
algebraic structure does not differ, we have Poincare
duality in the later case and degeneracy in the former.

Next step could be fine tuning of Poincare Duality definition. Since the 
review in \cite{asterisque}, Connes has preferred to take as primary 
definition the existence of an element $\beta \in KR_n(A^0\otimes A)$ such
that
$$ \beta \otimes_A \mu = 1_{A^0},  \mu \otimes_{A^0} \beta = 1_A $$ 
(where $\mu \in KR^n(A\otimes A^0)$ is got from our familiar Fredholm module). 

Perhaps we can not win enough space with this shift of mind:
It seems that in our case this definition implies the same isomorphism that the
intersection matrix reflects.   I have only seen an sketch of proof in
the PhD thesis of H. Emerson.  Also, it is remarked by Moscovici that
{\it rational Poincare Duality} based in the intersection form is
{\it weaker} than Poincare  duality based in the KK-bifunctor.

\section{Specific}

To further investigate what is happening inside, lets fix $n=6$. The
degeneracy space is spanned by the vectors $(1,1,0,-1-1,0)$ and
 $(0,1,1,0,-1,-1)$.

We could see it in this way: Suppose an element of $KK(\C,A)$ is given by
the projector $(1,1,0,0,0,0)$. The product with the intersection
matrix drives us to $(0,0,1,0,0,1)$. But take now the projector
$(0,0,0,1,1,0)$: it gives us the same element $(0,0,1,0,0,1)$ in 
$KK(A^0,\C)$. So given this element in the K-homology, we can not say
which one was the original: Duality is bounded to fail if we only
kept the information of the intersection matrix.

Technically, the map between Ramond-Ramond fields and D-Branes... er, 
between K-theory and K-homology, is given by Kasparov product through
the $\mu$ element. Regretly, the product is a very complicated operation,
described only by a few high level mathematicians. At least, I have 
been unable to find detailed examples in the (still too modern)
bibliography. 

Readers of this paper will know that an element of $KK^0(A,B)$ is
given by an A-B $C^*$ bimodule, graded, and an bounded operator
F.  We can get F from an unbounded Dirac operator,
D, with F=sign(D). And a third alternative exists, the use of
asymptotic morphism. This variety of viewpoints has its origin in
the difficulty to calculate the $F$ operator in the product. For a 
couple of algebras, $a \in KK(A,B)$ and $b \in KK(B,C)$ respectively, Kasparov
product $\otimes_B$ is built from the graded tensor product of subjacent
Hilbert spaces, with a new operator $F=F_a \oslash F_b$ which I can not
explain how to calculate. 
For the moment, I will try at the level of the graded tensor only. This is no
so bad because F is defined up homotopy and modulo compact operators. In fact,
the current formulation of classification theorems for finite spectral
triples \cite{kr1,PS} is able
to avoid any specific value of F.

Our algebra is $A=\C^6$. Let $a \in KK(\C,A)$ be based in the graded space
$H^+=(p_1 A\oplus p_2A), H^-=(p_4A\oplus p_5A)$. The Dirac element 
$\mu \in KK(A\otimes A^0,\C)$ is based in our old pal\cite{previous}, the 
space $\bigoplus E_{ij}$ with grading $\Gamma E_{ij}=$sign($q_{ij}$)$E_{ij}$.
And $a\otimes a^0 \in A\otimes A^0$ acts in this space
multiplying by $a_i a^0_j$.

With all this data, we only need to promote $KK(\C,A)$ to 
$KK(A^0, A\otimes A^0)$ with a trivial direct product and than compose
the Kasparov groups. The resulting space, in $KK(A^0,\C)$ has elements
$$H_{16}\oplus H_{21}\oplus H_{12}\oplus H_{23}\oplus H_{44} \oplus H_{55}
$$ in the positive grading, and elements
$$H_{11}\oplus H_{22}\oplus H_{43}\oplus H_{54}\oplus H_{45} \oplus H_{56}
$$ in the negative grading. 

Under the action of $A^0$, each part has the same number of vectors in the
positive and negative gradings. So the product with the K-theory will give
in every case an element homotopic to zero. 

In general, we see that a projector $p_i$ of positive grading will give three
vectors in the
product space: $H_{i,i-1}, H_{i,i+1}$ in the positive side, $H_{i,i}$ in the
negative. For the projectors described by $(1,1,0,0,0,0)$ and 
$(0,0,0,1,1,0) $ we have respectively Hilbert spaces
$$(H_{12}\oplus H_{16}\oplus H_{23}\oplus H_{21})\oplus (H_{11} \oplus H_{22})
$$ and 
$$(H_{45}\oplus H_{43}\oplus H_{54}\oplus H_{56}) \oplus (H_{44} \oplus H_{55})
$$  
 
If we only look at the action of $A^0$, we are still in the same 
situation: both spaces are the same, 
some information has been lost, and Kasparov product
$\beta \otimes_{A^0}$ is unable to bring us back to the original
element. 

But if we could remember the origin of each vector then
we could go back to the right one: the first projector gives the
space $H_{23}\oplus H_{16}$ and the second one gives 
      $H_{43}\oplus H_{56}$.  But if we want to distinguish between both
spaces, the $A^0$ action is not enough: we need to give again a role
to the Fredholm operator F (or to its "unbounded" counterpart, the
Dirac operator).

A way to restrict the homotopies of F could be searched in the ambiguity of the
Dirac operator. As we told in \cite{previous}, the limit procedure has
an angular freedom, we can define $m_{l-1\,l,ll}=(1/n) \sin \theta$,
$m_{ll,l+1\,l}=- (1/n) \cos\theta$ and we still get the same continuous
limit. If we study the index of $D$ as $\theta$ varies, Index Theorem works
and it is conserved. But we can notice that the symmetric derivative,
$\theta=\pi/4$, corresponds to a crossing of eigenvalues: just at that
point, a pair of eigenvectors of $D$ cross the zero. This could justify us
to avoid the symmetric point and to choose a nonsymmetric value for $\theta$.
Suppose you choose $\theta< \pi/4$. Then we have a distinction between the
Hilbert subspaces $H_{i-1,i}$ and $H_{i+1,i}$. This is more patent if 
we move to just the backward derivative, $\theta=0$, where one of the
subspaces is directly in the nullspace of $D$.

Really the extreme cases $0$ and $\pi/2$ amount to reduce the spectral
triple to a simpler one, so they are no so desiderable. It should be better
to kept just a slight asymmetry and to do the Kasparov product keeping
this Dirac operator all the way. Thus our conclusion is that more work
is needed in Kasparov products to define suitable dualities in 
finite algebras.

\section{Worth mentioning}

\begin{itemize}
\item
When taking the $n\to\infty$ limit, one needs more groups beyond the
$KK^0$. Fortunately they are directly produced by entering Clifford
algebras into play, and we get the needed generators in our limit.
(for the role of Clifford algebras in the real spectral triples, and all
the reduction and unreduction game, see the book \cite{FGV}).

\item
It could be interesting to build a relative KK doing contractions of two
points to one. It could be interesting also to see all the sequence of
discrete spaces as different formulations of discrete derivatives. And if
we are really, really interested in this play one should investigate if
the Hopf algebra of Connes-Moscovici has a role linking spaces of
different size, then aiming to go from a bare series to a renormalized one.

\item
If you remember the diagonalization procedure of the previous paper,
you could note that the vectors $E_{ij}+E_{ji}$ are in the kernel of the
Dirac operator, while
the vectors $E_{ij}-E_{ji}$ are just the piece one needs to counterweight the
$1/n$ divergence of D, then getting a finite contribution
for the one-dimensional derivative. Note also that the $E_{ii}$ vectors, for
the same reason, induce a divergent derivative, which is controlled because it
goes to an imaginary part: we have  "finite $+ i$ infinite".

\item
If the renormalization process forces the survival of some vestige of the
approximation procedure, then we will have a spectre of Connes-Lot particles
in the continuum. In some  sense this is a low-profile approach to the big 
project of studying quantum groups and general diffeomorfism. Note also
that a residual Dirac operator reflects, via Lichnerowitz, a curvature, and
then it is a way to proof that the spectrum of particles is bounded away
from zero, at least for the particular case coming from discretization
and renormalization. 

\item Of course the physical interest of getting asymmetry into the
Dirac operator is because we need diferent masses and mixes in the
generations of particles.

\end{itemize}


\begin{thebibliography}{20}


\bibitem{connes}
A. Connes, {\it Gravity coupled with matter and the foundation of
non commutative geometry}, hep-th/9603053

\bibitem{asterisque}
A. Connes, {\it Brisure de Symetrie Spontanee et Geometrie du Point de Vue
Spectral}, Asterisque, {\bf 241}, expose 816.

(Also included in vol. 2 of the "serie jornades cientifiques" del "Institut 
d'Estudis Catalans")
See also recent reviews in hep-th by this author.

\bibitem{FGV}
J.M. Gracia-Bondia, J.C. Varilly and H. Figueroa, {\it Elements of
Noncommutative Geometry}, Birkhauser, (2000).

\bibitem{kr1}
T. Krajewski, {\it Classification of Finite Spectral Triples},
arXiv:hep-th/9701081
\bibitem{PS}
M. Paschke, A. Sitarz, {\it Discrete spectral triples and their 
symmetries} arXiv:q-alg/9612029

\bibitem{previous}
A. Rivero, {\it Discrete spectral triples converging to Dirac operators}
 arXiv:math-ph/0203024



\end{thebibliography}
\end{document}